\documentclass[a4paper,UKenglish,cleveref, autoref, thm-restate]{lipics-v2021}

\pdfoutput=1 
\hideLIPIcs 

\usepackage{mathtools}
\usepackage{amssymb}

\usepackage[ruled]{algorithm}
\usepackage{algpseudocode}

\usepackage{tikz}
\usepackage{graphicx}

\theoremstyle{plain}
\newtheorem{maintheorem}{Theorem}

\newtheorem{property}[theorem]{Property}

\crefname{maintheorem}{Theorem}{Theorems}
\Crefname{maintheorem}{Theorem}{Theorems}

\DeclarePairedDelimiter{\ceil}{\lceil}{\rceil}

\newcommand{\OBPDD}{\textup{\textsc{OBP-DD}}}

\newcommand{\NextFit}{\textup{\textsc{Next-Fit}}}
\newcommand{\FirstFit}{\textup{\textsc{First-Fit}}}
\newcommand{\BestFit}{\textup{\textsc{Best-Fit}}}
\newcommand{\WorstFit}{\textup{\textsc{Worst-Fit}}}
\newcommand{\AnyFit}{\textup{\textsc{Any-Fit}}}
\newcommand{\FiveThirds}{\textup{\textsc{Five-Thirds}}}

\newcommand{\NF}{\mathrm{NF}}
\newcommand{\FF}{\mathrm{FF}}
\newcommand{\WF}{\mathrm{WF}}
\newcommand{\AF}{\mathcal{AF}}

\newcommand{\OPT}{\mathrm{OPT}}
\newcommand{\ALG}{\mathrm{ALG}}

\newcommand{\size}[1]{s(#1)}
\newcommand{\arr}[1]{\operatorname{arr}(#1)}
\newcommand{\dl}[1]{\operatorname{dl}(#1)}

\newcommand{\arrive}{\mathrm{arr}}
\newcommand{\close}{\mathrm{cl}}

\newcommand{\R}{\mathbb{R}}
\newcommand{\N}{\mathbb{N}}

\title{Online Bin Packing with Dispatch Deadlines} 

\author{Pei-Lin Chiang}{Department of Computer Science, National Tsing Hua University, Hsinchu, Taiwan}{ivylingchiang@gmail.com}{https://orcid.org/0009-0008-3435-2537}{}
\author{Wing-Kai Hon}{Department of Computer Science, National Tsing Hua University, Hsinchu, Taiwan}{wkhon@cs.nthu.edu.tw}{https://orcid.org/0000-0002-0570-2904}{Supported by NSTC, Taiwan, under Grant 113-2221-E-007-139-MY3.}
\author{Kazuo Iwama}{Department of Industrial Engineering and Engineering Management, National Tsing Hua University, Hsinchu, Taiwan}{iwama@kuis.kyoto-u.ac.jp}{}{Supported by NSTC, Taiwan, under Grants 114-2223-E-007-016 and NSTC 115-2223-E-007-012.}
\author{Li-Tzu Yang}{Department of Computer Science, National Tsing Hua University, Hsinchu, Taiwan}{breton0429@gmail.com}{https://orcid.org/0009-0007-3733-9605}{}

\authorrunning{P.-L. Chiang, W.-K. Hon, K. Iwama, and L.-T. Yang}

\Copyright{Pei-Lin Chiang, Wing-Kai Hon, Kazuo Iwama, and Li-Tzu Yang}



\acknowledgements{We thank the anonymous reviewers for their helpful comments on improving the presentation and for pointing us to additional related work. ChatGPT was used to assist with literature searches and with improving the authors' existing text.}

\nolinenumbers 

\begin{document}
\maketitle

\begin{abstract}
We study a generalization of \emph{online bin packing} by introducing dispatch deadlines. 
Items arrive over time in an online fashion. Each item has an arrival time, a size in $(0,1]$, and a deadline, all of which are revealed only when the item arrives.
The algorithm must immediately and irrevocably assign the item to a unit-capacity bin that has not yet been dispatched. Each bin must be dispatched no later than the earliest deadline of the items assigned to it, and once it is dispatched, it cannot be used for future items. The goal is to minimize the number of bins used.

Applications arise naturally in logistics. Additionally, the offline version of this problem has appeared under several names, including the \emph{clique partition problem in weighted interval graphs} and the \emph{BinSchedule} problem. This problem is also a special case of \emph{online bin packing with conflicts} and is closely related to online bin packing models with delay costs.

We study the competitive ratios of several classical online bin packing algorithms when applied to this problem, and show that a similar hierarchy of competitive ratios carries over. Our main technical results are that the competitive ratio of \FirstFit{} is exactly $16/7 \approx 2.28571$, and that no deterministic online algorithm has a competitive ratio less than $11/5 = 2.2$, showing that \FirstFit{} is quite close to best possible. In addition, \NextFit{} and \WorstFit{} have competitive ratio exactly~$3$, and every algorithm in the \AnyFit{} class has competitive ratio between $16/7$ and $3$.
\end{abstract}

\section{Introduction} 

\emph{Online bin packing} is one of the central problems in online algorithms. 
In this problem, items arrive one by one, each with a size in $(0,1]$. Upon arrival, an item must be immediately and irrevocably assigned to a unit-capacity bin, and the goal is to minimize the number of bins used. Finding optimal algorithms for this problem has been a long-standing line of work~\cite{balogh2015optimal,balogh2018upper,balogh2021lower}, and classical algorithms such as \FirstFit{} and \BestFit{} have been well-studied since the early work on bin packing~\cite{johnson1973phd,johnson1974worst,johnson1974fast}.

Classical online bin packing has no notion of \emph{deadlines}: once a bin is opened, it remains available forever. To add deadlines, at least two different approaches have appeared in the literature. 
One is the \emph{dynamic bin packing} problem, where items depart and are \emph{removed from their bins} at their deadlines, while the bins remain available for future items~\cite{coffman1983dynamic,chan2005dynamic,chan2009dynamic,wong2012dynamicLower}. 
The other is the \emph{bin dispatch} model: when an item's deadline is reached, the entire bin containing the item, which one may think of as a truck, must be dispatched and cannot be used for future items. 
This feature creates a tradeoff. Suppose that an item~$p$, with arrival time~$a$ and deadline~$d$, is relatively tight, meaning that $d-a$ is relatively small. Assigning~$p$ to an existing bin $B$ may save a bin now, but it may also force $B$ to be dispatched earlier, before later items that could have been packed into $B$ arrive.

While there has been considerable work on this model~\cite{berman2007packing} and its equivalent formulation~\cite{li2006minimum,chen2007some,myung2008clique}, all existing work studies the offline version. However, the tradeoff described above is also, and arguably even more, interesting in the online setting, where each packing decision must be made without knowing future items. In this paper, we study this online setting, which, to the best of our knowledge, has not appeared in the literature.
We call this problem \emph{Online Bin Packing with Dispatch Deadlines} (\OBPDD{}). An input is a finite sequence $I=(p_1,p_2,\ldots)$ of items, where each item is represented as a tuple $p_i = (a_i, s_i, d_i)$. Here, $a_i \in \R$ is the arrival time, $s_i \in (0,1]$ is the size, and $d_i \in \R$ is the deadline, with $a_i \le  d_i$. For convenience, we write $\arr{p_i} = a_i$, $\size{p_i} = s_i$, and $\dl{p_i} = d_i$. 
The input is revealed in an online fashion. When an item arrives, its arrival time, size, and deadline are revealed, and the algorithm must immediately and irrevocably assign it to a unit-capacity bin that is not yet dispatched. That is, for every bin $B$, we require $\sum_{p\in B}\size{p}\le 1$ and $\arr{p}\le \min_{q\in B}\dl{q}$ for every $p\in B$. The goal is to minimize the number of bins used.

The offline version has appeared under several names. If all items are known in advance, each item $p$ can be viewed as an interval $[\arr{p},\dl{p}]$ with weight $\size{p}$. A bin then corresponds to a set of intervals with a common intersection and total weight at most~$1$. Thus, the bin packing problem is equivalent to the \emph{clique partition problem in weighted interval graphs}~\cite{li2006minimum,chen2007some,myung2008clique}, where the goal is to partition the weighted intervals into the minimum number of bounded-weight cliques. This problem is strongly NP-hard and cannot be approximated within a factor of $3/2-\varepsilon$ in polynomial time for any $\varepsilon>0$ unless $\mathsf{P}=\mathsf{NP}$~\cite{li2006minimum,chen2007some}. A $2$-approximation algorithm for general weights and an exact algorithm for the identical-weight case were later given in~\cite{myung2008clique}. The same offline problem also appears as \emph{BinSchedule}, where items have weights, arrival times, and deadlines; Berman~et~al.~\cite{berman2007packing} gave a $3$-approximation algorithm in general and a $3/2$-approximation algorithm when the difference between the deadline and the arrival time is the same for all items.

As for the online version, note that \OBPDD{} is equivalent to classical online bin packing if all deadlines are sufficiently large.
Furthermore, \OBPDD{} is a special case of \emph{online bin packing with conflicts}. In this problem, a conflict graph specifies which pairs of items cannot share a bin, and each item reveals its size and its conflict edges to previous items upon arrival. 
Epstein and Levin~\cite{epstein2008conflicts} studied the problem for interval conflict graphs, giving a $4.7$-competitive algorithm and a lower bound of $155/36 \approx 4.30556$ for any online algorithm. 
In contrast, \OBPDD{} corresponds to the special case in which two items conflict exactly when their intervals $[\arr{p},\dl{p}]$ are disjoint (i.e., the conflict graph is co-interval), and the items arrive in the order of left endpoints.

Closely related to deadline models are delay models. In the \emph{online bin packing with delay and holding costs} problem~\cite{ahlroth2013delay}, each item $p$ has a nondecreasing continuous delay-cost function $D_p(t)$, where $t$ is the time elapsed from the arrival of $p$ until its bin is released; open bins may additionally incur holding costs. If discontinuous delay-cost functions taking values in the extended reals are allowed, then \OBPDD{} is equivalent to the special case with no holding costs and $D_p(t)=0$ for $t\leq \dl{p}-\arr{p}$ and $D_p(t)=\infty$ otherwise. Another delay model is the \emph{bin packing with delays} problem~\cite{azar2019clustering,epstein2021delays}, where the packing of an item may be delayed, incurring a nondecreasing continuous delay cost until the item is packed; whenever a set of items is packed into a bin, that bin is closed and cannot be used for future items. Similarly, deadlines can be represented using the same delay-cost function as above; however, since the packing of an item may be deferred, this is not equivalent to \OBPDD{}.

Real-world applications of \OBPDD{} arise in logistics, in particular in \emph{cross-docking}. In traditional logistics, storage and order picking are often the most costly warehouse operations. Cross-docking reduces these costs by consolidating freight from different origins into outbound vehicles with minimal handling and little or no storage between unloading and loading~\cite{belle2012cross}. In our abstraction, items are pieces of freight, bins are outbound trucks or containers, item sizes are capacity requirements, and deadlines are latest departure times.

\subparagraph*{Our Results.}
In this paper, we study the following natural question: how robust are classical online bin packing algorithms in the presence of dispatch deadlines? We address this question by directly applying these algorithms to \OBPDD{} and analyzing their performance.
Several classical algorithms are considered: \NextFit{}, \FirstFit{}, \WorstFit{}, and the \AnyFit{} class, which includes \FirstFit{} and \WorstFit{} but not \NextFit{}. In the classical setting, the following asymptotic competitive ratios are well known~\cite{johnson1973phd,johnson1974worst,johnson1974fast}: \NextFit{} has ratio~$2$; every \AnyFit{} algorithm has ratio at least $1.7$ and at most $2$; the lower bound $1.7$ is matched by \FirstFit{}, while the upper bound $2$ is matched by \WorstFit{}. 
We extend these algorithms to \OBPDD{} in the most direct way. The packing rule is unchanged; the only difference is that bins are no longer available once they are dispatched. Thus, when an item arrives, the algorithm applies its usual rule only to the bins that are still open, and, unless the algorithm explicitly closes a bin earlier, each bin is kept open until it is forced to be dispatched by a deadline.
Our results show that the same hierarchy of competitive ratios carries over from classical online bin packing to \OBPDD{}, but with larger ratios.

In~\cref{section:NF_AF_WF}, we first present rather straightforward results, which also illustrate fundamental differences between classical online bin packing and \OBPDD{}. We show that \NextFit{} has competitive ratio exactly $3$~(\cref{theorem:NF}), every algorithm in the \AnyFit{} class has competitive ratio at least $16/7$ and at most $3$~(\cref{theorem:AF}), and \WorstFit{} has competitive ratio exactly $3$~(\cref{theorem:WF}). The result for \WorstFit{} also shows that the upper bound of $3$ for the \AnyFit{} class is tight. 
In~\cref{section:FF}, we show that \FirstFit{} has competitive ratio exactly $16/7$~(\cref{theorem:FF}), which implies that the lower bound of $16/7$ for the \AnyFit{} class is tight.
In particular, \FirstFit{} has the best possible worst-case performance among all \AnyFit{} algorithms. This implies that \BestFit{}, another classical online bin packing algorithm often compared with \FirstFit{}, performs no better than \FirstFit{} in the worst case in \OBPDD{}, since \BestFit{} is in the \AnyFit{} class. 

The upper bound for \FirstFit{} is our main technical result. Its proof is based on an \emph{input-reduction} method. One challenge here is that \FirstFit{} does not use deadline values when deciding where to pack an item. In classical bin packing, the bins opened by \FirstFit{} remain available forever, so one can deduce bounds on the cumulative sizes of bins in their opening order; this is the basis of the standard area arguments used in bin packing analysis. In \OBPDD{}, however, bins may be dispatched and later bins may be opened in a highly interleaved way. This makes it difficult to deduce bounds on the cumulative sizes over a sequence of \FirstFit{} bins. To handle this, we give an input-reduction algorithm: given any input, it constructs another input on which \FirstFit{} uses the same number of bins, while the optimal offline solution uses no more bins, and the bins opened by \FirstFit{} are dispatched according to a more regular pattern. It then suffices to prove the upper bound based on inputs with this regular pattern instead.

Note that \FirstFit{} achieves competitive ratio $16/7$ without using the deadline information at all, which one might expect to severely lower the power of the algorithm. It turns out, however, that this is not the case, as $16/7 \approx 2.28571$ is quite close to the best possible ratio for \emph{any} deterministic online algorithm: in~\cref{section:GLB}, we show that no deterministic online algorithm can attain a competitive ratio less than $11/5 = 2.2$~(\cref{theorem:GLB}).
The trick is that our adversary can disrupt the online algorithm's decision-making ability: when a new item arrives, the online algorithm cannot tell whether the item's deadline is ``early'' or ``late''.
This lower bound also separates \OBPDD{} from classical online bin packing, for which a $5/3$-competitive algorithm is known~\cite{balogh2015optimal}. 

\section{Preliminaries}
\label{section:Preliminaries}

For an input $I = (p_1, p_2, \dots)$, a solution is a partition of the items into bins $B_1, B_2, \dots$ such that for every bin $B_j$, $\sum_{p \in B_j} \size{p} \le 1$ and no item $p \in B_j$ has $\arr{p} > \min_{q \in B_j} \dl{q}$. We call the partition $B_1, B_2, \dots$ a \emph{packing} of $I$. The cost of a packing is the total number of bins used. Our goal is to minimize the number of bins used.

We say that a bin is \emph{open} from the time it is opened until the time it is dispatched. For consistency with this terminology, we use the term \emph{close} instead of \emph{dispatch} from now on. The two terms have the same meaning.

Multiple items can have the same arrival time or deadline. Without loss of generality, we use the following convention, since it cannot increase the cost. If several items have the same arrival time, they are processed one by one in increasing order of their indices. If arrivals and deadlines occur at the same time, arrivals are processed first, and then bins with an item whose earliest deadline is that time are closed.

The cost of an algorithm on an input $I$ is the cost of the packing produced by the algorithm on $I$. We evaluate the performance of online algorithms via standard \emph{competitive analysis}. For an input $I$, let $\ALG(I)$ denote the cost of an online algorithm $\ALG$ on $I$, and $\OPT(I)$ denote the cost of an optimal offline algorithm $\OPT$ on $I$. The \emph{competitive ratio} of $\ALG$ is defined as $\sup_{I} \ALG(I)/\OPT(I)$.

Another common performance measure is the \emph{asymptotic competitive ratio}, defined as $\limsup_{n\to\infty}\sup_{I:\OPT(I)=n}\ALG(I)/\OPT(I)$. 
We remark that all competitive-ratio bounds in this paper also hold for the asymptotic competitive ratio. 
This is because all of our upper bounds hold without an additive term, and all of our lower bounds are constructed via sequences of inputs $I_N$ such that $\OPT(I_N)\to\infty$. 
For \NextFit{}, \WorstFit{}, and \FirstFit{}, the equivalence of the two measures can also
be shown directly by a simple input-repetition argument; see Appendix~\ref{appendix:competitive-ratio}.

\subparagraph*{Two Simple Bounds on OPT.}
Suppose $B_1, \dots, B_m$ is a packing of an input $I$. 
Then, let $\size{B_i}$ denote the total size of all items in a bin $B_i$, i.e., $\size{B_i} = \sum_{p \in B_i} \size{p}$, and $\size{I}$ the total size of all items in $I$, i.e., $\size{I} = \size{B_1} + \dots + \size{B_m}$.
Since each bin has capacity one, we obtain the following bound. 

\begin{lemma}
\label{lemma:size_bound}
If $B_1, \dots, B_m$ is a packing of an input $I$, then $\OPT(I) \ge \size{B_1} + \cdots + \size{B_m}$.
\end{lemma}

Two items $p$ and $q$ are said to be \emph{disjoint} if $[\arr{p}, \dl{p}] \cap [\arr{q}, \dl{q}] = \emptyset$ (i.e., $\dl{p} < \arr{q}$ or $\dl{q} < \arr{p}$). Notice that two disjoint items can never be packed in the same bin. This gives the following bound.

\begin{lemma}
\label{lemma:time_bound}
If there are $N$ pairwise disjoint items in an input $I$, then $\OPT(I)\ge N$.
\end{lemma}

\subparagraph*{Algorithms.}
We now define the classical online bin packing algorithms considered in this paper in the setting of \OBPDD{}. Unless an algorithm explicitly closes a bin earlier, a bin is kept open until the earliest deadline among its items is reached. Among the algorithms considered below, only \NextFit{} explicitly closes a bin early. 
\NextFit{} ($\NF$), which uses $O(1)$ space, maintains at most one open bin at any time; when an item $p$ arrives, $\NF$ packs $p$ into the open bin if one exists and $p$ fits in it; otherwise, $\NF$ closes the open bin if one exists and opens a new bin for $p$.
\AnyFit{} ($\AF$) is the class of online algorithms that never opens a new bin for an arriving item $p$ if $p$ fits in some currently open bin. 
\WorstFit{} ($\WF$) packs an arriving item $p$ into a currently open bin with largest remaining capacity among those in which $p$ fits; if no such bin exists, $\WF$ opens a new bin for $p$. 
\FirstFit{} ($\FF$) scans the currently open bins in the order in which they were opened, and packs an arriving item $p$ into the first bin in which $p$ fits; if no such bin exists, $\FF$ opens a new bin for~$p$.
Thus, $\FF \in \AF$ and $\WF \in \AF$, but $\NF \notin \AF$.

\section{Next-Fit, Any-Fit, and Worst-Fit}
\label{section:NF_AF_WF}

\subsection{Next-Fit}

\begin{maintheorem}
\label{theorem:NF}
The competitive ratio of \NextFit{} is exactly $3$.
\end{maintheorem}
\begin{proof}
We first show the lower bound. Let $N$ be a large integer, let $\varepsilon = 1 / (6N)$, and let $t_1 < u_1 < u_2 < \cdots < u_N < t_{\infty}$. Consider the input 
\[
I = (A_1, B_1, A_2, B_2, \dots, A_{2N}, B_{2N}, C_1, C_2, \dots, C_N),
\]
where for $i = 1, \dots, 2N$, $A_i = (t_1, 1/2 - \varepsilon, t_{\infty})$ and $B_i = (t_1, 3\varepsilon, t_1)$. Also, for $j = 1, \dots, N$, $C_j = (u_j, 2\varepsilon, u_j)$.
At time $t_1$, for each $i=1,\dots,2N$, $\NF$ packs $A_i$ and $B_i$ together in one bin; if $i < 2N$, it then opens a new bin for $A_{i+1}$, since $\size{A_i} + \size{B_i} + \size{A_{i+1}} > 1$. These $2N$ bins are closed at time $t_1$. Then, for each time $u_j$, $j=1,\dots,N$, $\NF$ opens one bin for~$C_j$, which is closed immediately at time $u_j$. Thus, $\NF(I) = 2N+N = 3N$.
On the other hand, $\OPT$ can pack all items $B_1,\dots,B_{2N}$ together in one bin at time $t_1$. For each time $u_j$, $j=1,\dots,N$, $\OPT$ can pack $A_{2j-1}$, $A_{2j}$, and $C_j$ together in one bin. Thus, $\OPT(I) \le 1+N$, resulting in a lower bound of $3$ for the competitive ratio of $\NF$ as $N \to \infty$.

We next show the upper bound. Consider an input $I$. Let $B_1, \dots, B_{\NF(I)}$ be the $\NF$ packing of $I$, indexed by the order in which they are opened. We call $B_i$ an \emph{urgent} bin if $\size{B_i} + \size{B_{i+1}} \le 1$ or $i = \NF(I)$. Let $K$ be the total number of urgent bins.
Consider an urgent bin $B_i$ with $i<\NF(I)$, and let $p$ be the first item packed into $B_{i+1}$. Then $B_i$ must have been closed before $p$ arrived. Indeed, if $B_i$ were still open when $p$ arrived, then $\NF$ would have packed $p$ into $B_i$, since $\size{B_i} + \size{p} \le \size{B_i} + \size{B_{i+1}} \le 1$. Thus, $B_i$ contains an item whose deadline is before $\arr{p}$. Therefore, each urgent bin can be associated with an item such that these associated items are pairwise disjoint. By~\cref{lemma:time_bound}, $\OPT(I) \ge K$.
On the other hand, for every non-urgent bin $B_i$, we have $i<\NF(I)$ and $\size{B_i}+\size{B_{i+1}}>1$. There are $\NF(I)-K$ non-urgent bins. Summing this inequality over all non-urgent bins, each bin size appears in at most two terms, and hence $\size{I} \ge (\NF(I)-K)/2$. By~\cref{lemma:size_bound}, $\OPT(I) \ge (\NF(I)-K)/2$.
Combining the two bounds, we obtain $\NF(I) \le 3 \, \OPT(I)$.
\end{proof}

The lower bound input above also illustrates the difference between classical online bin packing and \OBPDD{}. The last items $C_1,\dots,C_N$ play the role of a ``garbage collection'' step: they are small items that $\OPT$ can use to fill the remaining space next to pairs of $A_i$-items, but that $\NF$ is forced to pack separately because its earlier bins have already been closed. In the classical setting, equivalently when all items have deadline $t_\infty$, the last $\NF$ bin would remain open and would have enough space to contain all items $C_1,\dots,C_N$. Thus, the same input would make $\NF$ use only $2N$ bins in the classical setting, rather than $3N$.

\subsection{The Any-Fit Class}

\begin{maintheorem}
\label{theorem:AF}
The competitive ratio of any algorithm in the \AnyFit{} class is at least $16/7$ and at most~$3$.
\end{maintheorem}
\begin{proof}
Let $\ALG \in \AF$. 
We first show the lower bound.
Let $N$ be a large integer divisible by $3$, let $\varepsilon = 1/(9N)$, and let $t_1 < \cdots < t_N < u_1 < \cdots < u_{4N/3} < t_\infty$. Consider the input
\[
\begin{aligned}
I = (&A_1, B_1, C_1, D_1, E_1, F_1, G_1, H_1, J_1, \dots, A_N, B_N, C_N, D_N, E_N, F_N, G_N, H_N, J_N,\\
&K_1, \dots, K_{4N/3}),
\end{aligned}
\]
where for $i = 1, \dots, N$, $A_i = (t_i, 3\varepsilon, t_i)$, $B_i = (t_i, 1/3 - 9\varepsilon, t_\infty)$, $C_i = (t_i, 1/3 - \varepsilon, t_\infty)$, $D_i = (t_i, 9\varepsilon, t_\infty)$, $E_i = G_i = J_i = (t_i, 1/3 - \varepsilon, t_i)$, and $F_i = H_i = (t_i, 1/3 + 3\varepsilon, t_\infty)$. 
Also, for $j = 1, \dots, 4N/3$, $K_j = (u_j, 3\varepsilon, u_j)$.
By the \AnyFit{} rule and the choice of item sizes, for each time $t_i$, $i = 1, \dots, N$, $\ALG$ packs items $A_i, B_i, C_i, D_i$ in one bin, $E_i, F_i$ in one bin, $G_i, H_i$ in one bin, and $J_i$ in one bin; these four bins are closed at time $t_i$. Then, for each time $u_j$, $j=1,\dots,4N/3$, $\ALG$ opens one bin for $K_j$, which is closed immediately at time $u_j$. Thus, $\ALG(I) = 4N + 4N/3 = 16N/3$. 
On the other hand, for each time~$t_i$, $i=1,\dots,N$, $\OPT$ can pack $A_i, E_i, G_i, J_i$ in one bin. For each time $u_j$, $j=1,\dots,N$, $\OPT$ can pack $B_j, F_j, H_j, K_j$ in one bin. For each time $u_j$, $j=N+1,\dots,4N/3$, $\OPT$ can pack $C_{3(j-N)-2}, C_{3(j-N)-1}, C_{3(j-N)}, K_j$ in one bin. Finally, $\OPT$ can pack $D_1, \dots, D_N$ in one~bin. Thus, $\OPT(I) \le N + 4N/3 + 1 = 7N/3 + 1$, resulting in a lower bound of $16/7$ for the competitive ratio of $\ALG$ as $N \to \infty$. The two packings are illustrated in~\cref{figure:AF_lower}.

\begin{figure}[t]
    \centering
    \resizebox{\linewidth}{!}{%
        \input{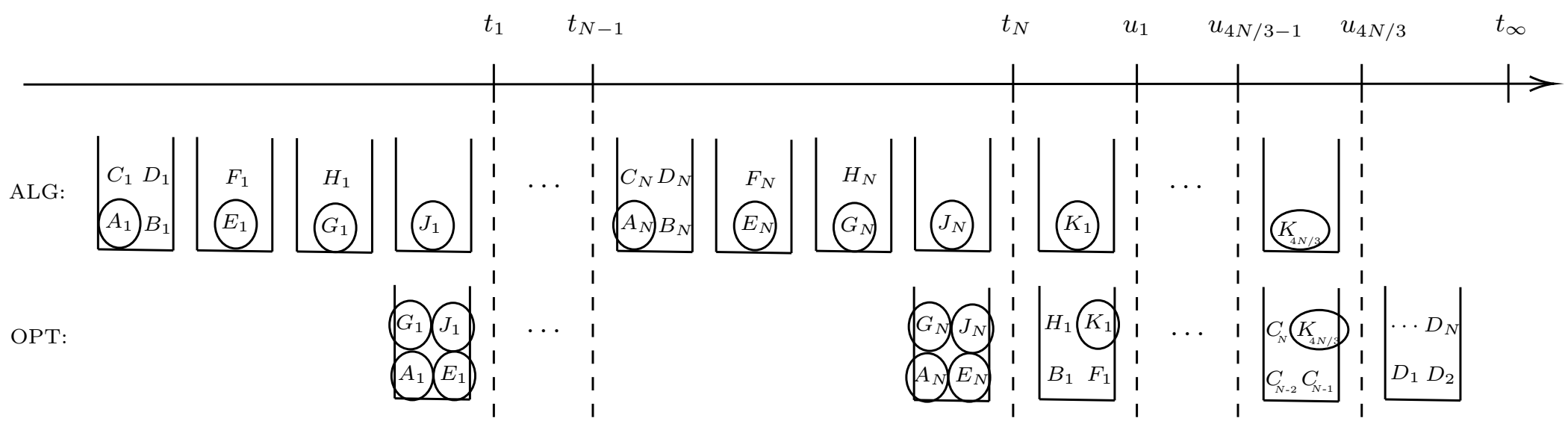}
    }
    \caption{$\ALG$ and $\OPT$ packings for the $\AF$ lower-bound construction. Item sizes are not drawn to scale. Each bin is closed at the nearest dashed line to its right. The circled items are the items whose deadlines cause the corresponding bins to be closed at that time. At each time $t_i$, each of the first three $\ALG$ bins has size $\approx 2/3$, while the fourth has size $\approx 1/3$; at each time $u_j$, the $\ALG$ bin has size $\approx 0$. In contrast, all $\OPT$ bins are full.}
    \label{figure:AF_lower}
\end{figure}

We next show the upper bound. Consider an input $I$. Let $S$ and $L$ be the numbers of bins $B$ with $\size{B} \leq 1/2$ and $\size{B} > 1/2$, respectively, in the $\ALG$ packing of $I$. By the $\AF$ rule, no two of these $S$ bins can be open at the same time; otherwise, the first item of the later-opened bin would fit into the earlier one. Thus, each of these $S$ bins can be associated with an item (the item whose deadline forces it to close) such that these associated items are pairwise disjoint. By~\cref{lemma:time_bound}, $\OPT(I) \geq S$. On the other hand, by~\cref{lemma:size_bound}, we have $\OPT(I) \geq L/2$. Combining the two bounds, we obtain $\ALG(I) = S+L \leq 3 \, \OPT(I)$.
\end{proof}

For completeness, we remark that the upper bound could also be proved by combining~\cref{theorem:NF} with the stronger property that $\ALG(I) \le \NF(I)$ for every input $I$ and $\ALG \in \AF$. 
This property is well-known in classical online bin packing~\cite{boyar2012FF}, and the proof carries over almost identically to \OBPDD{}, with the additional case that an $\NF$ bin may have been closed by a deadline.
We give the proof in Appendix~\ref{appendix:AF}.

\subsection{Worst-Fit}

\begin{maintheorem}
\label{theorem:WF}
The competitive ratio of \WorstFit{} is exactly $3$.
\end{maintheorem}
\begin{proof}
Since $\WF \in \AF$, by~\cref{theorem:AF}, the competitive ratio of $\WF$ is at most $3$. Notice that the same lower bound input from~\cref{theorem:NF} applies to $\WF$ as well ($\WF$ produces the exact same packing as $\NF$ for the input). Thus, the competitive ratio of $\WF$ is at least $3$.
\end{proof}

\section{First-Fit}
\label{section:FF}

\begin{maintheorem}
\label{theorem:FF}
The competitive ratio of \FirstFit{} is exactly $16/7$.
\end{maintheorem}

By~\cref{theorem:AF}, the competitive ratio of $\FF$ is at least $16/7$. It remains to prove that the competitive ratio of $\FF$ is at most $16/7$, which we do in the rest of this section.
At a high level, our proof consists of two steps. First, we give an algorithm that, given any input $I$, constructs another input $I'$ in a special form such that $\FF(I)/\OPT(I) \le \FF(I')/\OPT(I')$. Second, we show that $\FF(I')/\OPT(I') \le 16/7$ for any input $I'$ of that special form. Let us first define the special form as follows. Notice that the second property of this form implies that all bins opened at time~$t_i$ are closed at $t_i$ as well.

\begin{definition}
\label{definition:canonical_form}
We say that an input $I$ is in \emph{canonical form} if there exist times $t_1 < \cdots < t_N < t_\infty$, for some $N \in \N$, such that the $\FF$ packing of $I$ satisfies the following properties.
\begin{enumerate}
    \item At least one bin is opened at time $t_i$, for every $i = 1, \dots, N$, and no bin is opened at any other time.
    \item For every $i = 1, \dots, N$, every bin opened at time $t_i$ contains exactly one item with deadline~$t_i$, called its \emph{trigger item}, and all other items in the bin have deadline $t_\infty$.
    \item For every $i = 1, \dots, N$, the trigger items of all bins opened at time $t_i$ have total size at most $1$.
\end{enumerate}
\end{definition}

We now give the reduction algorithm; see~\cref{algorithm:reduction}. 
Informally, the main idea of the reduction is to take ``snapshots'' of the $\FF$ packing. Notice that at any time $t$, if we select any subsequence of the currently open $\FF$ bins and form a new input consisting only of the items currently in these bins, in their original arrival order, then running $\FF$ from scratch on this new input produces exactly these bins. Thus, to obtain a canonical input, we can partition the $\FF$ bins into groups, take a snapshot of each group at a separate time, and give one item in each bin of the snapshot an early deadline so that the whole group closes before the next snapshot. 
Two questions remain: how to choose the trigger items so that their total size is at most $1$, and how to make sure that the new input, consisting of these snapshotted items, is no harder for $\OPT$.
The key is to take snapshots based on $\OPT$'s packing.
Whenever $\OPT$ closes a bin, we take a snapshot of the not-yet-snapshotted $\FF$ bins that contain items from this $\OPT$ bin, and use one of these items in each $\FF$ bin as its trigger item. 
Since all these trigger items were packed in one $\OPT$ bin, they have total size at most $1$.
Moreover, the original $\OPT$ packing gives a feasible packing of the new input.

\begin{algorithm}[t]
\caption{Reduction to canonical form}
\label{algorithm:reduction}
\begin{algorithmic}[1]
\Require An input $I$
\Ensure An input $I'$ in canonical form

\State Assume that all arrival times and deadlines in $I$ are distinct. \label{line:assumption}
\State Run $\FF$ on $I$, and fix an $\OPT$ packing $B^*_1,\ldots,B^*_m$ of $I$, indexed by increasing order of their closing times.
\State Let $t_i$ be the closing time of $B^*_i$, for $i=1,\dots,m$, and let $t_\infty > t_m$.

\State Mark all bins in the $\FF$ packing of $I$ as unmarked, and set $I' \leftarrow I$.

\For{$i \leftarrow 1,\dots,m$} \label{line:for_loop}
    \State List the items in $B^*_i$ as $p_1,\dots,p_n$.
    \For{$j \leftarrow 1,\dots,n$}
        \State Let $B(p_j)$ be the bin containing $p_j$ in the $\FF$ packing of $I$.
        \If{$B(p_j)$ is unmarked}
            \State Mark $B(p_j)$.
            \State Remove all items in $B(p_j)$ whose arrival times are after $t_i$ from $I'$.
            \State Change the arrival times of the remaining items in $B(p_j)$ to $t_i$ in $I'$.
            \State Change the deadline of $p_j$ to $t_i$ in $I'$.
            \State Change the deadlines of the other remaining items in $B(p_j)$ to $t_\infty$ in $I'$.
        \EndIf
    \EndFor
\EndFor

\State Sort the items in $I'$ by their new arrival times, breaking ties by their original arrival order in $I$.
\State \Return the resulting input $I'$.
\end{algorithmic}
\end{algorithm}

The assumption on Line~\ref{line:assumption} is without loss of generality, 
by an event-ordering argument; see Appendix~\ref{appendix:distinct-times}.
We now prove the correctness of~\cref{algorithm:reduction}.

\begin{lemma}
\label{lemma:reduction}
For every input $I$,~\cref{algorithm:reduction} returns an input $I'$ in canonical form such that $\FF(I')=\FF(I)$ and $\OPT(I')\le \OPT(I)$.
\end{lemma}
\begin{proof}
Let us continue to use the notation in~\cref{algorithm:reduction}.
For convenience, discard every index $i\in\{1,\dots,m\}$ such that no bin in the $\FF$ packing of $I$ is marked during the processing of $B_i^*$ on the $i$-th iteration of the for-loop on Line~\ref{line:for_loop}. Let $N\le m$ be the number of remaining indices, and relabel the remaining pairs $(t_i,B_i^*)$, preserving their original order, as $(t_1,B_1^*),\dots,(t_N,B_N^*)$.
For every item $p$ of $I$ that is not removed from $I'$, let $p'$ denote the corresponding item in $I'$. Thus, $\size{p'}=\size{p}$. For a set $S$ of items of $I$, let $S'=\{p' \mid p\in S \text{ and } p \text{ is not removed from } I'\}$.

After the execution of~\cref{algorithm:reduction}, every bin in the $\FF$ packing of $I$ is marked exactly once. This is because every such bin contains an item belonging to some $B_i^*$, so it is eventually marked; moreover, once a bin is marked, it is never marked again.
Hence, every item in $I$ is either removed from $I'$, or changed to have arrival time $t_i$ and deadline either $t_i$ or $t_\infty$, for some $i\in\{1,\dots,N\}$. Therefore, all items in $I'$ arrive at one of the times $t_1,\dots,t_N$, and so in the $\FF$ packing of $I'$, no bin is opened at any other time.

For each $i=1,\dots,N$, let $\mathcal M_i$ be the set of bins in the $\FF$ packing of $I$ that are first marked during the processing of $B_i^*$. By the relabeling above, $\mathcal M_i$ is nonempty for every $i$. Write $\mathcal M_i=\{B_{i,1},\dots,B_{i,k_i}\}$, where the bins are ordered by their opening order in the $\FF$ packing of $I$. For each $r=1,\dots,k_i$, define $S_{i,r}=\{p\in B_{i,r} \mid \arr{p}\le t_i\}$.
These are exactly the items of $B_{i,r}$ that remain in $I'$ after $B_{i,r}$ is marked. Thus, the items arriving at time $t_i$ in $I'$ are exactly the items in $S'_{i,1},\dots,S'_{i,k_i}$. Moreover, by the final sorting step of~\cref{algorithm:reduction}, their arrival order at $t_i$ is the same as the arrival order of their corresponding items in $I$.

Note that for every $i$ and $r$, the items removed from $B_{i,r}$ are exactly the items that arrive after $t_i$ in $I$. Thus, when comparing the packing of the items in $S'_{i,r}$ at time $t_i$ in $I'$ with the packing of their corresponding items in $I$, the removed items were not present yet in $I$, and so they do not affect any ``fit'' or ``non-fit'' decision of $\FF$.

\begin{claim}
\label{claim:each_block}
Let $i\in\{1,\dots,N\}$. In the $\FF$ packing of $I'$, if all bins opened before time $t_i$ are closed before time $t_i$, then the bins opened at time $t_i$ are exactly $S'_{i,1},\dots,S'_{i,k_i}$.
\end{claim}
\begin{claimproof}
For convenience, in this claim, an \emph{$\FF[I']$-bin} means a bin in the $\FF$ packing of $I'$.
We prove this claim by induction over the items arriving at time $t_i$. 
For the base case, consider the first item arriving at time $t_i$ (i.e., the item corresponding to the bin-opening item of $B_{i,1}$). By the assumption of the claim, all bins opened before time $t_i$ are already closed, so $\FF$ opens a new bin for this item. We call this bin the $\FF[I']$-bin corresponding to $B_{i,1}$.

For the inductive step, consider an item $p'\in S'_{i,r}$, for some $r \in \{1,\dots,k_i\}$, and suppose that every earlier item $q'\in S'_{i,\ell}$ has already been packed into the $\FF[I']$-bin corresponding to $B_{i,\ell}$, for every $\ell=1,\dots,k_i$. We want to show that $p'$ is also packed into the $\FF[I']$-bin corresponding to $B_{i,r}$.

First, consider any $\ell<r$. The first item of $B_{i,\ell}$ belongs to $S_{i,\ell}$ and arrives before $p$ in $I$, so by the induction hypothesis, the $\FF[I']$-bin corresponding to $B_{i,\ell}$ has already been opened in $I'$. Since $B_{i,\ell}$ is first marked during the processing of $B_i^*$, no item in $B_{i,\ell}$ has deadline before $t_i$; otherwise, the $\OPT$ bin containing that item would close before $t_i$, and $B_{i,\ell}$ would have been marked earlier. Thus, $B_{i,\ell}$ is still open when $p$ arrives in $I$. Since $\FF$ packs $p$ into $B_{i,r}$ in $I$, the item $p$ does not fit into $B_{i,\ell}$. Moreover, by the induction hypothesis, when $p'$ arrives in $I'$, the $\FF[I']$-bin corresponding to $B_{i,\ell}$ contains exactly the items $q'$ such that $q\in B_{i,\ell}$ and $q$ arrives before $p$ in $I$. Hence this bin has the same total size as $B_{i,\ell}$ had when $p$ arrived in $I$. Therefore, $p'$ does not fit into the $\FF[I']$-bin corresponding to $B_{i,\ell}$ in $I'$.

Now consider $B_{i,r}$. If some earlier item of $S'_{i,r}$ has already arrived in $I'$, then by the induction hypothesis, these earlier items have been packed into the $\FF[I']$-bin corresponding to $B_{i,r}$. Moreover, this bin contains exactly the items $q'$ such that $q\in B_{i,r}$ and $q$ arrives before $p$ in $I$. Hence it has the same total size as $B_{i,r}$ had when $p$ arrived in $I$, so $p'$ fits in this bin. Since $p'$ does not fit into any earlier bin, $\FF$ packs $p'$ into this bin.

It remains to consider the case where no earlier item of $S'_{i,r}$ has arrived in $I'$ yet. 
This means $p'$ is the first item of $S'_{i,r}$ to arrive, and so $p$ is the bin-opening item of $B_{i,r}$.
Therefore, no $\FF[I']$-bin corresponding to $B_{i,\ell}$ with $\ell>r$ has been opened yet when $p'$ arrives. Since $p'$ does not fit into any earlier bin, $\FF$ opens a new bin (i.e., the $\FF[I']$-bin corresponding to $B_{i,r}$) for $p'$.

By induction over the items arriving at time $t_i$, the bins opened at time $t_i$ are exactly $S'_{i,1},\dots,S'_{i,k_i}$.
\end{claimproof}

\begin{claim}
\label{claim:reduction_bins}
In the $\FF$ packing of $I'$, for every $i=1,\dots,N$, the bins opened at time $t_i$ are exactly $S'_{i,1},\dots,S'_{i,k_i}$.
\end{claim}
\begin{claimproof}
By~\cref{claim:each_block}, it suffices to show that, for every $i=1,\dots,N$, all bins opened before time $t_i$ are closed before time $t_i$.
We prove this statement by induction on $i$.
For $i=1$, no bin is opened before time $t_1$, and so the statement is vacuous.
For the inductive step, suppose that all bins opened before time $t_i$ are closed before time $t_i$. By~\cref{claim:each_block}, the bins opened at time $t_i$ are exactly $S'_{i,1},\dots,S'_{i,k_i}$. For each $r=1,\dots,k_i$, $S'_{i,r}$ contains the item that caused $B_{i,r}$ to be marked, and this item has deadline $t_i$ in $I'$. Hence, every bin opened at time $t_i$ is closed at time $t_i$. Therefore, if $i<N$, all bins opened before time $t_{i+1}$ are closed before time $t_{i+1}$, and so the statement holds for $i+1$.
\end{claimproof}

Now, consider the $\FF$ packing of $I'$. By~\cref{claim:reduction_bins}, for every $i=1,\dots,N$, the bins opened at time $t_i$ are exactly $S'_{i,1},\dots,S'_{i,k_i}$. Since $\mathcal M_i$ is nonempty for every $i$, at least one bin is opened at each time $t_i$. We have already shown that no bin is opened at any other time.
Moreover, for each $r=1,\dots,k_i$, exactly one item in $S'_{i,r}$ has deadline $t_i$, namely the item that caused $B_{i,r}$ to be marked, and all other items in $S'_{i,r}$ have deadline $t_\infty$. 
Finally, all items with deadline $t_i$ at time $t_i$ were packed in $B_i^*$, so their total size is at most $1$. Therefore, the three properties in~\cref{definition:canonical_form} are satisfied, and so $I'$ is in canonical form.

Recall that every bin in the $\FF$ packing of $I$ is marked exactly once. Therefore, the sets $\mathcal M_1,\dots,\mathcal M_N$ partition the bins in the $\FF$ packing of $I$, and hence $\FF(I)=\sum_{i=1}^N k_i$.
On the other hand, by~\cref{claim:reduction_bins}, the $\FF$ packing of $I'$ opens exactly $k_i$ bins at each time $t_i$, and no bin is opened at any other time. Thus, $\FF(I')=\sum_{i=1}^N k_i=\FF(I)$.

Finally, we show that the original $\OPT$ packing of $I$ is a packing of $I'$, except that some items are removed from the bins.
In this paragraph, we use the original indexing $B_1^*,\dots,B_m^*$ from~\cref{algorithm:reduction}.
Consider a bin $B^*_i$, for some $i \in \{1,\dots,m\}$, and an item $p_j \in B^*_i$. 
If $p_j$ is removed in $I'$, then we are done, so suppose not. 
If $B(p_j)$ is first marked during the processing of $B^*_i$, then $\arr{p_j}$ is changed to $t_i$ and $\dl{p_j}$ is changed to either $t_i$ or $t_\infty$ in~$I'$. Since $t_i < t_\infty$, packing $p_j$ in the same $\OPT$ bin is still valid in $I'$.
If $B(p_j)$ is already marked during the processing of $B^*_{i'}$, for some $i' < i$, then $\arr{p_j}$ is changed to $t_{i'}$ and $\dl{p_j}$ is changed to either $t_{i'}$ or $t_\infty$. Notice that the former is impossible since $p_j$ is not in $B^*_{i'}$, and so $\dl{p_j}$ is changed to $t_\infty$. Since $t_{i'} < t_i < t_\infty$, packing $p_j$ in the same $\OPT$ bin is still valid in $I'$. 
Thus, $\OPT(I') \le \OPT(I)$.
\end{proof}

By the above lemma, to prove $\FF(I) / \OPT(I) \le 16/7$ for every input $I$, it is sufficient to prove it for those inputs $I$ that are in canonical form. 
Before we proceed, we state a simple algebraic property that will be used repeatedly.

\begin{property}
\label{property:fraction}
Let $A,B,C,D\in \R$ with $A+C>0$, $A+D>0$, $A\le B$, and $C\le D$. Then $(B+D)/(A+D)\le (B+C)/(A+C)$.
\end{property}

We now prove the $16/7$ upper bound. Let $I$ be an input in canonical form with times $t_1 < \cdots < t_N < t_\infty$. For every $i = 1, \dots, N$, let $K_i$ be the number of $\FF$ bins opened and closed at time $t_i$, and let $S_i$ be the total size of all items in these $K_i$ bins.
Fix a \emph{lazy} $\OPT$ packing of $I$, where a lazy packing means every bin is kept open until the deadline of some item in the bin forces $\OPT$ to close the bin. Thus, $\OPT$ bins are closed only at times $t_1, \dots, t_N$ and possibly $t_\infty$. For every $i = 1,\dots,N$, let $K^*_i$ be the number of $\OPT$ bins closed at time $t_i$, and let $S_i^*$ be the total size of all items in these $K_i^*$ bins.
The following lemma shows that it suffices to work on a single time $t_i$.

\begin{lemma}
\label{lemma:group_suffices}
If for every $i=1,\dots,N$, $(K_i+S_i-S_i^*)/(K_i^*+S_i-S_i^*)\le 16/7$, then $\FF(I) / \OPT(I) \le 16/7$.
\end{lemma}
\begin{proof}
Let $K^*_\infty$ be the number of $\OPT$ bins closed at time $t_\infty$. Notice that $K^*_\infty\ge \sum_{i=1}^{N} S_i-\sum_{i=1}^{N} S_i^* \ge 0$ by~\cref{lemma:size_bound}, since the remaining items not packed in the first $\sum_{i=1}^{N} K_i^*$ OPT bins have total size $\sum_{i=1}^{N} S_i-\sum_{i=1}^{N} S_i^*$. Thus, we have
\[
\frac{\FF(I)}{\OPT(I)}
=
\frac{\sum_{i=1}^{N}K_i}{\sum_{i=1}^{N}K_i^*+K^*_\infty}
\le
\frac{\sum_{i=1}^{N}K_i+K^*_\infty}{\sum_{i=1}^{N}K_i^*+K^*_\infty}
\le
\frac{\sum_{i=1}^{N}K_i+\sum_{i=1}^{N} S_i-\sum_{i=1}^{N} S_i^*}{\sum_{i=1}^{N}K_i^*+\sum_{i=1}^{N} S_i-\sum_{i=1}^{N} S_i^*},
\]
where the last inequality follows from~\cref{property:fraction}.

By the assumption of the lemma, for every $i=1,\dots,N$, $K_i+S_i-S_i^*\le (16/7)(K_i^*+S_i-S_i^*)$. Summing over all indices gives $\sum_{i=1}^{N}K_i+\sum_{i=1}^{N} S_i-\sum_{i=1}^{N} S_i^*\le (16/7)(\sum_{i=1}^{N}K_i^*+\sum_{i=1}^{N} S_i-\sum_{i=1}^{N} S_i^*)$. Therefore $\FF(I)/\OPT(I)\le 16/7$.
\end{proof}

We now show that the sufficient condition in the above lemma indeed holds.

\begin{lemma}
\label{lemma:group_ratio}
For every $i=1,\dots,N$, $(K_i+S_i-S_i^*)/(K_i^*+S_i-S_i^*)\le 16/7$.
\end{lemma}
\begin{proof}
For convenience, let us redefine the notation to focus on a single time $t_i$: let $K = K_i$, $K^* = K^*_i$, $S = S_i$, and $S^* = S^*_i$.
Let $B_1,\dots,B_K$ be the $\FF$ bins opened and closed at time $t_i$, indexed by the order in which they are opened.
Notice that we have $S^*\le K^*$ and $K^*\ge 1$. Hence, if $K^* > K$ or $K = 1$, then the ratio is at most $1$ already. Let us assume $K^* \le K$ and $K > 1$. We derive two bounds for the ratio.

\begin{claim}
\label{claim:bound_1}
For every $K > 1$, $(K+S-S^*)/(K^*+S-S^*)\le (5K-4)/(2K-1)$.
\end{claim}
\begin{claimproof}
We first show that $S\ge (2K-1)/3$. Notice that for every $i<K$, the bin $B_i$ contains at least two items. This is because otherwise $B_i$ contains only its trigger item, so the trigger items of $B_i$ and $B_K$ have total size greater than $1$, a contradiction to~\cref{definition:canonical_form}.
Furthermore, among $B_1,\dots,B_{K-1}$, at most one bin has size less than $2/3$. This is because if $B_i$ is the first such bin, then every item in a later bin has size greater than $1/3$, and so every non-last later bin has size greater than $2/3$. Let $B_i$ be such a bin if it exists, and otherwise choose any bin among $B_1,\dots,B_{K-1}$. Since $B_K$ is opened after $B_i$, we have $\size{B_i}+\size{B_K}>1$. Hence $S\ge (2/3)(K-2)+1=(2K-1)/3$.

Therefore, we have $S-S^*\ge (2K-1)/3-K^*$. By~\cref{property:fraction},
\[
    \frac{K+S-S^*}{K^*+S-S^*}
    \le
    \frac{K+\frac{2K-1}{3}-K^*}{\frac{2K-1}{3}}
    \le
    \frac{K+\frac{2K-1}{3}-1}{\frac{2K-1}{3}}
    =
    \frac{5K-4}{2K-1}.
\]
\end{claimproof}

\begin{claim}
\label{claim:bound_2}
For every $K>1$, $(K+S-S^*)/(K^*+S-S^*)\le 2+1/(K-1)$.
\end{claim}
\begin{claimproof}
For $i=2,\dots,K$, let $\alpha_i=\max_{j<i}(1-\size{B_j})$. Then, every item in $B_i$ has size greater than $\alpha_i$. On the other hand, since the trigger items fit in one bin by~\cref{definition:canonical_form}, $\sum_{i=2}^K\alpha_i < 1$. Notice that $\alpha_2\le\alpha_3\le\cdots\le\alpha_K$, and $\size{B_i}\ge 1-\alpha_{i+1}$ for every $i<K$. Thus,
\[
    S
    \ge \sum_{i=1}^{K-1}(1-\alpha_{i+1})+\alpha_K
    \ge K-1-\sum_{i=2}^K\alpha_i+\frac{1}{K-1}\sum_{i=2}^K\alpha_i
    \ge K-2+\frac{1}{K-1}.
\]
Therefore, we have $S-S^*\ge K-2+1/(K-1)-K^*$. By~\cref{property:fraction},
\[
    \frac{K+S-S^*}{K^*+S-S^*}
    \le
    \frac{K+K-2+\frac{1}{K-1}-K^*}{K-2+\frac{1}{K-1}}
    \le
    \frac{K+K-2+\frac{1}{K-1}-1}{K-2+\frac{1}{K-1}}
    \le
    2+\frac{1}{K-1}.
\]
\end{claimproof}

\Cref{claim:bound_1} gives the desired bound for $1 < K\le 4$, and~\cref{claim:bound_2} gives it for $K\ge 5$.
\end{proof}

In summary,~\cref{lemma:reduction} shows that for every input $I$, there exists an input $I'$ in canonical form such that $\FF(I)/\OPT(I) \le \FF(I')/\OPT(I')$;~\cref{lemma:group_suffices} and~\cref{lemma:group_ratio} show that for every input $I'$ in canonical form, $\FF(I') / \OPT(I') \le 16/7$.
We have thus proven that the competitive ratio of $\FF$ is at most $16/7$.

\section{General Lower Bound}
\label{section:GLB}
In this section, we show a lower bound of $11/5$ for the competitive ratio of any deterministic online algorithm.
Let us first give the intuition behind our adversary.
Fix any deterministic online algorithm $\ALG$. Similar to the $\AF$ lower bound, the adversary tries to force $\ALG$ to open bins in groups. In each group, every bin opened by $\ALG$ contains one item with an early deadline and other items with late deadlines. The early-deadline items force the bins of $\ALG$ to close, while the late-deadline items can later be garbage-collected and packed more efficiently by $\OPT$.
The difficulty is that $\ALG$ may try to pack the early-deadline items together. To prevent this, the adversary does not decide in advance whether a newly released item is early or late. Instead, it maintains three variables $t_\arrive < t_\close < t_\infty$. Every new item arrives at time $t_\arrive$ and is given a deadline in $(t_\close, t_\infty)$. If $\ALG$ opens a new bin for this item, the adversary declares it to be an early item by updating $t_\close$ to its deadline. If $\ALG$ packs the item into an old bin, the adversary declares it to be a late item by updating $t_\infty$ to its deadline.
Thus, after each update, every bin-opening item has deadline no later than the current~$t_\close$, while every non-bin-opening item has deadline no earlier than the current~$t_\infty$. Since the updates only increase $t_\close$ or decrease $t_\infty$, and always preserve $t_\close < t_\infty$, these two properties hold throughout the construction. Once the adversary has forced the desired number of bins, it waits until time $t_\close$, so all bins are closed.
It then chooses new values
of $t_\arrive$ and $t_\close$ such that $\text{old }t_\close < t_\arrive < t_\close < t_\infty$, and starts the next group.

Let us now give the algorithm for the adversary; see~\cref{algorithm:GLB_adversary}.

\begin{algorithm}[t]
\caption{Adversary for the $11/5$ general lower bound.}
\label{algorithm:GLB_adversary}
\begin{algorithmic}[1]
\Require A deterministic online algorithm $\ALG$.

\State Let $N$ be a large power of two and $\varepsilon = 1 / (12 N \cdot 11^{\lg{N}})$.
\State Initialize $t_\arrive$, $t_\close$, and $t_\infty$ to be any real numbers such that $t_\arrive < t_\close < t_\infty$.
\For{$i \leftarrow 1,\dots,\lg{N}+1$}
    \State Let $N_i = N/2^{i-1}$ and $\varepsilon_i = 11^{i-1}\varepsilon$.
    \For{$j \leftarrow 1,\dots,N_i$}
        \State Release items for iteration $(i,j)$ according to the procedure described below. \label{line:GLB_iteration}
        \State $t_\arrive \leftarrow t_\close + (t_\infty - t_\close)/3$.
        \State $t_\close \leftarrow t_\close + 2(t_\infty - t_\close)/3$.
    \EndFor
\EndFor

\State Let $M = \ceil{N/3} + (N/2 + N/4 + N/8 + \dots + 1)$ and $\delta = (t_\infty - t_\arrive)/M$.
\For{$m \leftarrow 1, \dots, M$}
    \State Release an item $x_m = (t_\arrive + m\delta, \varepsilon, t_\arrive + m\delta)$.
\EndFor

\end{algorithmic}
\end{algorithm}

\subparagraph*{Item-release procedure.}
Whenever we say that the adversary releases an item of size $s$ by the \emph{update rule}, we mean the following. Let $t=(t_\close+t_\infty)/2$. Release the item $p=(t_\arrive,s,t)$. If $\ALG$ opens a new bin for $p$, then update $t_\close\leftarrow t$; otherwise, update $t_\infty\leftarrow t$.

At the beginning of iteration $(i,j)$, no $\ALG$ bin is open at time $t_\arrive$. The adversary releases items by the update rule, with item sizes chosen according to the following decision tree.
First release two items $A$ and $B$ of size $1/3-2\varepsilon_i$. The item $A$ must open a new bin. If $B$ opens a new bin, release an item $C$ of size $1/3-\varepsilon_i$. If $C$ opens a new bin, the procedure ends. Otherwise, release an item $D$ of size $1/3+4\varepsilon_i$. If $D$ opens a new bin, the procedure ends. Otherwise, release an item $E$ of size $1/3+4\varepsilon_i$, which must open a new bin, and the procedure ends.
If instead $B$ is packed with $A$, release an item $C'$ of size $6\varepsilon_i$. If $C'$ is packed with $A$ and $B$, release two items $D'$ and $E'$ of size $1/3-\varepsilon_i$ as follows. The item $D'$ must open a new bin. If $E'$ opens a new bin, the procedure ends. Otherwise, release an item $F$ of size $1/3+3\varepsilon_i$, which must open a new bin, and the procedure ends. If $C'$ opens a new bin, release an item $D''$ of size $1/3+5\varepsilon_i$. If $D''$ opens a new bin, the procedure ends. Otherwise, release an item $E''$ of size $2/3-10\varepsilon_i$, which must open a new bin, and the procedure ends.
The nine possible ways in which $\ALG$ can pack the above items are summarized below, where the bins are listed in their opening order:
\begin{center}
\begin{tabular}{c|c|c|c}
case & $\ALG$ packing & bin-opening items & remaining items \\
\hline
1 & $\{A\},\{B\},\{C\}$ & $A,B,C$ & -- \\
2 & $\{A,C\},\{B\},\{D\}$ & $A,B,D$ & $C$ \\
3 & $\{A\},\{B,C\},\{D\}$ & $A,B,D$ & $C$ \\
4 & $\{A,C\},\{B,D\},\{E\}$ & $A,B,E$ & $C,D$ \\
5 & $\{A,D\},\{B,C\},\{E\}$ & $A,B,E$ & $C,D$ \\
6 & $\{A,B,C'\},\{D'\},\{E'\}$ & $A,D',E'$ & $B,C'$ \\
7 & $\{A,B,C'\},\{D',E'\},\{F\}$ & $A,D',F$ & $B,C',E'$ \\
8 & $\{A,B\},\{C'\},\{D''\}$ & $A,C',D''$ & $B$ \\
9 & $\{A,B\},\{C',D''\},\{E''\}$ & $A,C',E''$ & $B,D''$
\end{tabular}
\end{center}

For convenience, let us use the notation $(s)$-item and $(\le s)$-item to denote items with size exactly $s$ and at most $s$, respectively.
The following lemma follows directly from the update rule and the nine cases listed above.

\begin{lemma}
\label{lemma:GLB_iteration}
For every iteration $(i,j)$, the procedure described above releases items satisfying the following properties, where $t_\close$ and $t_\infty$ denote their values at the end of the procedure.
\begin{enumerate}
    \item All these items arrive at $t_\arrive$, and $\ALG$ opens three bins for them.
    \item The three bin-opening items have deadlines no later than $t_\close$ and total size at most $1$.
    \item The remaining items consist of at most one $(\le 1/3-\varepsilon_i)$-item, at most one $(\le 1/3+5\varepsilon_i)$-item, and at most one $(6\varepsilon_i)$-item. All such items have deadline no earlier than $t_\infty$.
\end{enumerate}
\end{lemma}

\begin{maintheorem}
\label{theorem:GLB}
No deterministic online algorithm has a competitive ratio less than $11/5$.
\end{maintheorem}
\begin{proof}
Let $\ALG$ be any deterministic online algorithm, and consider the input $I$ generated by~\cref{algorithm:GLB_adversary}.
By~\cref{lemma:GLB_iteration}, after each iteration $(i,j)$, $\ALG$ opens three bins and all three bins are closed before the next iteration starts. Therefore all the $(i,j)$ iterations contribute $3\sum_{i=1}^{\lg{N}+1} N_i = 3(N+N/2+\cdots+1)$ bins to $\ALG$. Moreover, since the final $M$ items are pairwise disjoint, $\ALG$ opens one bin for each of them. Thus, $\ALG(I) \ge 3(N+N/2+\cdots+1) + \ceil{N/3} + (N/2+N/4+\cdots+1) \ge 22N/3-O(1)$.

We now describe an $\OPT$ packing. For each iteration $(i,j)$, pack the three bin-opening items of the iteration in one bin. These bins contribute $\sum_{i=1}^{\lg{N}+1}N_i = N+N/2+\cdots+1$ bins.
It remains to pack the remaining items from the iterations, together with the final items $x_1,\dots,x_M$. For the items from iteration $i=1$, there are at most $N$ $(\le 1/3-\varepsilon_1)$-items. Pack three such items with one final item $x_m$ in one bin. This uses at most $\ceil{N/3}$ bins.
Next, for each $i=1,\ldots,\lg{N}$, pack two $(\le 1/3+5\varepsilon_i)$-items from iteration $i$, one $(\le 1/3-\varepsilon_{i+1})$-item from iteration $i+1$, and one final item $x_m$ in one bin. Since $\varepsilon_{i+1}\ge11\varepsilon_i$, each such bin has total size at most $1$. This uses at most $N_i/2=N/2^i$ bins for iteration $i$, and hence at most $N/2+N/4+\cdots+1$ bins in total. The remaining $(\le 1/3+5\varepsilon_{\lg{N}+1})$-item from the last iteration, if it exists, is packed in one additional bin.
Finally, for each iteration $i$, there are at most $N_i$ remaining $(6\varepsilon_i)$-items. Their total size is at most $\sum_{i=1}^{\lg{N}+1} 6N_i\varepsilon_i \le (N+N/2+\cdots+1)\cdot6\cdot 11^{\lg{N}}\varepsilon \le 12N\cdot 11^{\lg{N}}\varepsilon \le 1$, and so all these items fit in one additional bin.
Therefore, $\OPT(I) \le (N+N/2+\cdots+1)+\ceil{N/3}+(N/2+N/4+\cdots+1)+2 \le 10N/3+O(1)$.

Thus, $\ALG(I) \ge 22N/3-O(1)$ and $\OPT(I)\le 10N/3+O(1)$, resulting in a lower bound of $11/5$ for the competitive ratio of $\ALG$ as $N \to \infty$.
\end{proof}

\section{Further Directions}

The most immediate question is to narrow the gap between $11/5$ and $16/7$. 
Our results suggest that classical online bin packing algorithms with good competitive ratios may continue to perform well in \OBPDD{}. A natural next step is to analyze the \FiveThirds{} algorithm~\cite{balogh2015optimal}, which is optimal in classical online bin packing.
Another direction is to ask whether deadline information can be used to improve the ratio at all.
On the lower-bound side, one could try to strengthen~\cref{lemma:GLB_iteration} so that the algorithm is forced to open four bins, rather than three. This appears difficult, since the item sizes would need to satisfy many inequalities.

In online bin packing, \BestFit{} performs on par with \FirstFit{}, while in dynamic bin packing it performs worse than \FirstFit{}~\cite{chan2005dynamic}. 
It would be interesting to determine which behavior occurs in \OBPDD{}.

Another question is to identify restrictions on arrival times and deadlines that can make the problem easier. Recall that \OBPDD{} is equivalent to classical online bin packing when all deadlines are sufficiently large, or when all items have the same arrival time and deadline. What happens between these two extremes? For example, what is the competitive ratio when arrival times and deadlines take only two values, or when each item has a tight deadline, such as $\dl{p}\le \arr{p}+1$ in a discrete-time model?

\bibliography{references}

\appendix

\section{Asymptotic Competitive Ratio}
\label{appendix:competitive-ratio}

For algorithms such as $\NF$, $\WF$, and $\FF$, the equivalence of the two measures (competitive ratio and asymptotic competitive ratio) can also be proved by a simple input-repetition argument.
The main idea is that every (small) input can be repeated multiple times without changing the algorithm's behavior on each copy.

Specifically, let $I$ be an input, and let $\Delta > \max_{p\in I}\dl{p}-\min_{p\in I}\arr{p}$.
For $n \in \N$, define $I^n=I\cdot I_{\Delta}\cdot I_{2\Delta}\cdots I_{(n-1)\Delta}$, where $\cdot$ is the sequence concatenation operator and $I_{i\Delta}$ is obtained from $I$ by increasing the arrival time and deadline of every item by $i\Delta$.

Consider $\ALG\in\{\NF,\WF,\FF\}$.
Notice that the following property holds: $\ALG(I^n) = n \, \ALG(I)$.
This is because, by the choice of $\Delta$, all bins used for one copy are closed before the next copy begins, and $\ALG$ produces the same packing for each copy.
However, this property need not hold for an arbitrary $\AF$ algorithm.
For example, consider an algorithm that uses $\WF$ during the time interval $[0,1]$ and $\FF$ everywhere else.

We now show that this property implies equivalence of the two ratios.
Let $R$ and $R^\infty$ denote the competitive ratio and asymptotic competitive ratio of $\ALG$, respectively.
By definition, $R^\infty \le R$.
For the other direction, fix any input $I$ and consider $I^n$ as defined above.
Notice that $\OPT(I^n) = n \, \OPT(I)$ and, by the property, $\ALG(I^n) = n \, \ALG(I)$.
Hence, $R^\infty\ge \lim_{n\to\infty}\ALG(I^n)/\OPT(I^n)=\ALG(I)/\OPT(I)$.
Since this holds for every input $I$, we have $R^\infty\ge R$.
Thus, $R^\infty=R$.

\section{Comparison of Any-Fit and Next-Fit}
\label{appendix:AF}

We prove the property that $\NF$ uses at least as many bins as any $\AF$ algorithm on every input.

\begin{lemma}
For every input $I$ and $\ALG \in \AF$, $\ALG(I) \le \NF(I)$.
\end{lemma}
\begin{proof}
Let $B_1,\dots,B_{\ALG(I)}$ be the $\ALG$ packing of $I$, indexed by the order in which the bins are opened. For each $i=1,\dots,\ALG(I)$, let $p_i$ be the first item packed by $\ALG$ into $B_i$, and let $b(i)$ be the index of the bin in the $\NF$ packing that contains $p_i$.
We prove by induction on $i$ that $b(i)\ge i$ for every $i$. Clearly $b(1)=1$. Suppose $b(t)\ge t$ for some $t$.
If $\NF$ has not already opened bin $b(t)+1$ before processing $p_{t+1}$, then it has packed all items between $p_t$ and $p_{t+1}$ (including $p_t$ but excluding $p_{t+1}$) into bin $b(t)$.
If bin $b(t)$ of $\NF$ is closed before $p_{t+1}$ arrives, then $\NF$ opens bin $b(t)+1$ for $p_{t+1}$. 
Thus, suppose bin $b(t)$ of $\NF$ is still open when $p_{t+1}$ arrives. When $\ALG$ opens $B_{t+1}$ for $p_{t+1}$, the items currently packed in $B_t$ are a subset of the items packed by $\NF$ in bin $b(t)$. Since bin $b(t)$ of $\NF$ is still open, $B_t$ is also still open. Moreover, since $\ALG$ opens a new bin for $p_{t+1}$, the item $p_{t+1}$ does not fit in $B_t$. Therefore, $p_{t+1}$ cannot fit in bin $b(t)$ of $\NF$ either, and so $\NF$ packs $p_{t+1}$ into a later bin. 
In all cases, $b(t+1) \ge b(t)+1 \ge t+1$. By induction, $b(i)\ge i$ for every $i$. Thus, $\NF(I) \ge \ALG(I)$.
\end{proof}

\section{Justification of the Distinct-Time Assumption}
\label{appendix:distinct-times}

We justify the assumption on Line~\ref{line:assumption} of Algorithm~\ref{algorithm:reduction} that all arrival times and deadlines are distinct. We show that any input can be perturbed to satisfy this assumption without changing the packing produced by $\FF$ and the optimal cost.

Since the input is finite, we may choose $\varepsilon>0$ smaller than half the minimum distance between any two distinct event times (where an event is an arrival or a deadline). Consider any time $t$ at which multiple events occur. We perturb the arrivals at $t$ to distinct times in $(t-\varepsilon,t)$, while preserving their original processing order, and perturb the deadlines at $t$ to distinct times in $(t,t+\varepsilon)$. 

At every item arrival, the same previous items have arrived and the same bins have already closed as in the original input. Thus, by induction over the item arrivals, $\FF$ produces the same packing.
Moreover, two items are disjoint after the perturbation if and only if they are disjoint before the perturbation. Therefore, the set of all feasible packings is unchanged, and so is the optimal cost.

\end{document}